# Crystal Optical Properties of Inhomogeneous Plates and the Problems of Polarization Tomography of Photoelastic Materials

[1]Kushnir O., [1]Nek P., [2]Vlokh R., [3]Kukhtarev N.

[1]Electronics Department, Lviv National University, 107 Tarnavski St., 79017 Lviv, Ukraine
e-mail: o_s_kushnir@electronics.wups.lviv.ua
[2]Institute of Physical Optics, 23 Dragomanov St., 79005 Lviv, Ukraine
[3]Physics Department, Alabama A&M University, Normal, AL 35762
e-mail: nkukhtarev@aamu.edu

**Abstract**

Using the Jones matrix formalism, crystal optical properties of inhomogeneous material consisting of a pile of weakly birefringent plates are analysed in relation to the cell model adopted in polarization tomography of 3D dielectric tensor field in photoelastic media. It is shown that the material manifests in general an "apparent" optical gyration caused by different orientations of the plates. Relations between the polarimetric parameters and the dielectric tensor components are ascertained for the case of weak optical anisotropy.

## Introduction

Photoelasticity has been remaining one of extensively explored topics within the physical optics during the last decades (see, e.g., [1-4]). Its problems become enormously complicated if external (or internal) mechanical stresses are arbitrarily and inhomogeneously distributed inside a sample under test, thus transforming initially isotropic, macroscopically uniform material into anisotropic and homogeneous one. This field is often referred to as a 3D photoelasticity. In case of weakly anisotropic, weakly inhomogeneous materials possessing none or slight absorption, the stressed state manifests itself mainly in a notable effect on the polarization of probing light wave. Then the 3D spatial distribution of stresses (or, equivalently, that of dielectric parameters at the optical frequencies) might be clarified, using the methods of polarization optical tomography of 3D tensor fields (see [1,3]). In spite of essential current progress, those methods have not yet succeeded in solving the problem in its most general form.

Considerable analytical difficulties of the tensor field tomography have stipulated a number of approximate approaches, e.g., a "discretely inhomogeneous" (or "cell") model of the stressed material (see [2,5]), in which the sample is divided into many identical uniform cells. For a given propagation direction, the light beam probes a pile of anisotropic cells characterised by different parameters. This situation reminds a canonical problem of crystal optics, propagation of light through the so-called composite retardation plates [6,7]. Hence, elaboration of the latter would contribute to a better understanding of principles, techniques and difficulties of 3D tensor field tomography. It is the more so as a majority of tomographic problems are not solved analytically and, sometimes, even the appropriate physical interpretation is still lacking.

In this work we develop a matrix approach to the cell tomographic model and give a relevant interpretation of its results. For more clarity, though without any harm for generality, in many cases we shall restrict ourselves to the simplest case of the cell model, a cubic sample consisting of $N^3$ cubic cells, with $N = 2$. To make our consideration self-contained, we also remind and thoroughly substantiate some concepts related to the subject.

**Jones matrix description of the cell model**

Let both the optical anisotropy defined by a total birefringence and the optical inhomogeneity associated with relative spatial changes in the dielectric characteristics be small. Then the polarization effect of the sample can be correctly described in terms of Jones matrix (JM) approach [6-8] (see also [9]). According to simple symmetry analysis, the stressed state of initially isotropic particular cell, involving all nonzero stress tensor components, results in optical biaxiality of the cell. In the Cartesian reference frame, of which $z$ axis coincides with the propagation direction and $x$ and $y$ axes lie in perpendicular plane, the corresponding JM may be written as (see [10,11])

$$\mathbf{t} = F\mathbf{t}_n = F \begin{pmatrix} \cos\frac{\Delta}{2} - i\frac{l}{\Delta}\sin\frac{\Delta}{2} & \frac{-c+ip}{\Delta}\sin\frac{\Delta}{2} \\ \frac{c+ip}{\Delta}\sin\frac{\Delta}{2} & \cos\frac{\Delta}{2} + i\frac{l}{\Delta}\sin\frac{\Delta}{2} \end{pmatrix},$$

$$\Delta = \sqrt{l^2 + p^2 + c^2},$$
$$l = \Delta_l + i\delta_l, \quad p = \Delta_p + i\delta_p, \quad c = \Delta_c + i\delta_c$$
(1)

Here $\mathbf{t}_n$ represents the normalised [8] part of $\mathbf{t}$, $\Delta_l$, $\Delta_p$ and $\Delta_c$ the phase retardations respectively due to linear, "diagonal" linear and circular birefringences, $\delta_l$, $\delta_p$ and $\delta_c$ the corresponding parameters of linear, "diagonal" linear and circular dichroisms (see [10]), and $F$ the "isotropic" factor ($F = \exp(i\eta)$, $\eta = (2\pi d/\lambda)(\bar{n} + i\bar{\kappa})$, with $d$ being the thickness of the cell, $\lambda$ the light wavelength in vacuum and $\bar{n}$ and $\bar{\kappa}$ the mean refractive and extinction indices). Let us consider a lossless optical material, whose absorption coefficient and dichroism are absent ($\delta_l$, $\delta_p$, $\delta_c$, $\bar{\kappa} = 0$). Furthermore, the material is initially isotropic and centrosymmetric and so lacks any circular birefringence. Nor can this property be induced by the stresses, owing to a piezogyration effect [12]. As a consequence, we may put $\Delta_c = 0$. Eq. (1) then becomes

$$\mathbf{t}_n = \begin{pmatrix} \cos\frac{\Delta}{2} - i\cos 2\varphi\sin\frac{\Delta}{2} & i\sin 2\varphi\sin\frac{\Delta}{2} \\ i\sin 2\varphi\sin\frac{\Delta}{2} & \cos\frac{\Delta}{2} + i\cos 2\varphi\sin\frac{\Delta}{2} \end{pmatrix}, \quad (2)$$

with $\tan 2\varphi = \Delta_p / \Delta_l$ and $\Delta = \sqrt{\Delta_l^2 + \Delta_p^2}$, or $\mathbf{t}_n = \mathbf{R}(-\varphi)\mathbf{t}_{LB}\mathbf{R}(\varphi)$, where $\mathbf{R}(\varphi)$ is the rotation matrix by the angle $\varphi$ and

$$\mathbf{t}_{LB} = \begin{pmatrix} \exp(-i\Delta/2) & 0 \\ 0 & \exp(i\Delta/2) \end{pmatrix}$$ 

the JM of linearly birefringent crystal defined in the principal coordinate system. In other words, optical behaviour of a transparent, homogeneously stressed cell may be reduced to that of optically uniaxial crystal plate, whose principal axes are rotated by $\varphi$ with respect to $x$ and $y$ axes.

The optical properties of the entire sample along the propagation direction are determined by the JM $\mathbf{T} = \mathbf{t}_N...\mathbf{t}_1$. At this point a number of important questions arise, concerned with the quantity of independent parameters that might be derived, their physical meaning and relevant polarimetric techniques. From the fundamental point of view, answering the first question would need involving the "transverse" dielectric tensor $\mathbf{\varepsilon}_T$, which finally determines the JM (see [7,9]). Since the material is transparent, the only restriction imposed on $\mathbf{\varepsilon}_T$ is its Hermitian character ($\varepsilon_{T,ij} = \varepsilon_{T,ji}*$ [13]). Taking both real and imaginary parts into account ($\varepsilon_{T,ij} = \varepsilon'_{T,ij} + i\varepsilon''_{T,ij}$), we have in general four independent real parameters $\varepsilon_{T,11}$, $\varepsilon_{T,22}$, $\varepsilon'_{T,12}$ and $\varepsilon''_{T,11}$. In a quite equivalent manner, these are the diagonal components $\varepsilon_{11}$ and $\varepsilon_{22}$ defining the refractive indices of linearly polarised normal waves, the orientation angle $\varphi$ of optical indicatrix and the gyration tensor component $g_{33}$ [7,12,13].

On the other hand, the corresponding JM should be unitary [8,14], thus keeping invariant the norm of Jones vectors and so the light intensity. Sufficient conditions of the unitarity ($T_{n,22} = T_{n,11}*$, $T_{n,21} = -T_{n,12}*$ and $\det \mathbf{T}_n = 1$ [8,10]) give rise to five conditions for the real and imaginary parts of $T_{n,ij}$ and so remain only three independent numbers that fully determine this JM. Turning to the physical parameters, we may choose these as a polarization azimuth $\varphi$ of one of the normal waves (coincident with the orientation angle of the optical indicatrix), total phase retardation $\Delta$ and an ellipticity angle $\beta$ of the normal wave polarization. Another choice has been suggested in [14] (a primary and secondary characteristic azimuths and a characteristic retardation), which represents a set of experimental polarimetric quantities rather than the parameters strictly related to the physical properties. Finally, one more number, $\eta$, appears if one remembers that the JM $\mathbf{T} = \exp(i\eta)\mathbf{T}_n$ remains unitary, provided that $\mathbf{T}_n$ is. This fourth parameter is equivalent to the mean refractive index $\bar{n}$ or the corresponding mean dielectric constant ($\bar{n} = \sqrt{\varepsilon}$), which in fact describes the properties of the non-stressed material, too.

If the optical material is spatially inhomogeneous along the propagation direction, as in our case, the "inverse problem" of crystal optics (i.e., ascertaining its properties and characteristics on the basis of experimental polarization optical data) becomes cumbersome. We adopt here the approach consisting in introduction of "effective" homogeneous material, whose polarization-optical response is the same as that of the sample under consideration (see [15-18]). As follows from the said above, our "effective" material may in general manifest both linear ($\Delta n_l$, the phase retardation $\Delta_l$) and circular ($\Delta n_c$, the phase retardation $\Delta_c$) birefringences, as well as rotation of the optical indicatrix axes ($\varphi$). This can be expressed via the JM

$$\mathbf{T} = e^{-i\frac{2\pi d}{\lambda}\bar{n}} \mathbf{R}(-\varphi) \begin{pmatrix} \cos\frac{\Delta}{2} - i\cos 2\beta \sin\frac{\Delta}{2} & -\sin 2\beta \sin\frac{\Delta}{2} \\ \sin 2\beta \sin\frac{\Delta}{2} & \cos\frac{\Delta}{2} + i\cos 2\beta \sin\frac{\Delta}{2} \end{pmatrix} \mathbf{R}(\varphi), \qquad (3)$$

with $\Delta = \sqrt{\Delta_l^2 + \Delta_c^2}$, $\tan 2\beta = \Delta n_c / \Delta n_l$, $\Delta_l = (2\pi d / \lambda)\Delta n_l$ and $\Delta_c = (2\pi d / \lambda)\Delta n_c$.

Let us compare Eq. (3) with the JM $\mathbf{T} = \mathbf{t}_N...\mathbf{t}_1$. Using the notations $T_{ij} = T'_{ij} + iT''_{ij}$, we get

$$\tan 2\varphi = -\frac{T''_{12}}{T''_{11}}, \quad \sin 2\beta = \mp \frac{T'_{12}}{\sqrt{1-T'^2_{11}}}, \quad \cos\frac{\Delta}{2} = T'_{11}. \quad (4)$$

This is readily specified for the simplest case of $N = 2$:

$$\tan 2\varphi = \frac{\sin 2\varphi_1 \sin\frac{\Delta_1}{2}\cos\frac{\Delta_2}{2} + \sin 2\varphi_2 \cos\frac{\Delta_1}{2}\sin\frac{\Delta_2}{2}}{\cos 2\varphi_1 \sin\frac{\Delta_1}{2}\cos\frac{\Delta_2}{2} + \cos 2\varphi_2 \cos\frac{\Delta_1}{2}\sin\frac{\Delta_2}{2}}, \quad (5)$$

$$\sin 2\beta = \pm \frac{\sin 2(\varphi_2 - \varphi_1)\sin\frac{\Delta_1}{2}\sin\frac{\Delta_2}{2}}{\sqrt{1 - \left(\cos\frac{\Delta_1}{2}\cos\frac{\Delta_2}{2} - \cos 2(\varphi_2-\varphi_1)\sin\frac{\Delta_1}{2}\sin\frac{\Delta_2}{2}\right)^2}}, \quad (6)$$

$$\cos\frac{\Delta}{2} = \cos\frac{\Delta_1}{2}\cos\frac{\Delta_2}{2} - \cos 2(\varphi_2 - \varphi_1)\sin\frac{\Delta_1}{2}\sin\frac{\Delta_2}{2}, \quad (7)$$

where the total sample thickness is equal to $d = d_1 + d_2$ ($d_1 = d_2$) and the numbering ("1" or "2") corresponds to the queue, in which light beam is passing through the cells.

### Interpretation of optical effects in inhomogeneous material

A useful analysis of optical effects possible in such inhomogeneous material, representing a composite retardation palate, is now straightforward. The inhomogeneity manifests itself in different linear birefringences ($\Delta_1 \neq \Delta_2$, due to $\Delta n_{l1} \neq \Delta n_{l2}$) and different optical indicatrix orientations ($\varphi_1 \neq \varphi_2$) in the cells. The latter seems to be the most important factor, since it produces an optical activity ($\Delta_c, \beta \neq 0$), in spite of the fact that the constituent cells are not active ($\Delta_{c1}, \Delta_{c2} = 0$). This "apparent" optical gyration has nothing to do with the common physical mechanism linked with a so-called spatial dispersion in macroscopically uniform dielectric media (see [13,19]) and is completely due to the inhomogeneity. As with all the other following conclusions, we have checked this one to be also true of more complicated composite retardation plates ($N > 2$). So, we have $\beta = 0$ for $N = 3$ only if $\varphi_1 = \varphi_2 = \varphi_3$. One can prove that the gyration described by Eq. (6) has a close relation to the properties demonstrated by interference Šolc filters [7] at the working wavelength, for which the sequence of differently oriented linearly birefringent plates has a clearly nonzero transmittance, when placed between crossed polarizers (see the analysis [16]). Moreover, the results concerned with a novel gyration mechanism have been proven [16] to remain valid for a more realistic case of "continuously" (in particular, sinusoidally) inhomogeneous materials. It is interesting that the specific condition $\cos\Delta_1 \cos\Delta_2 - \sin 2(\varphi_2 - \varphi_1)\sin\Delta_1\sin\Delta_2 = 1$ (see Eq. (6)), fulfilled at $\varphi_2 - \varphi_1 = \pi/4$ and $\Delta_1 = \Delta_2 = \pi$ (half-wave constituent plates), results in $\beta = \pm\pi/4$, i.e. the sample exhibits a pure optical rotation of light polarization plane by the angle $\pi/2$, whereas the (normally dominating) linear birefringence in the cells is completely compensated.

Let us also mark that, under the reversal of propagation direction (the wave vector $\mathbf{k} \to -\mathbf{k}$ and "1"$\leftrightarrow$"2" – see [15]), we have $\mathbf{T_k} \neq \mathbf{T_{-k}}$ (because the matrices $\mathbf{t}_1,...,\mathbf{t}_N$ do not commute with each other) and alteration of the $\beta$ sign, while the parameters $\varphi$ and $\Delta$ are kept invariable (see, e.g., Eqs. (5)–(7)). This agrees well with what is known of the ordinary optical activity behaviour under the operation $\mathbf{k} \to -\mathbf{k}$ [12,13,19]. Moreover, probing of photoelastic sample in the opposite direction would in fact provide no extra experimental information.

Besides of the "effective" gyration, the inhomogeneous sample shows some other unusual properties. Eqs. (5) and (7) testify nontrivial general relations $\Delta \neq \Delta_1 + \Delta_2$ and $\varphi \neq \varphi_1, \varphi_2$ for the total retardation and the optical indicatrix angle, respectively. When, in particular, $\varphi_2 - \varphi_1 = \pm \pi / 4$, the composite plate consisting of two half-wave plates turns out, rather curiously, to be also half-wave ($\Delta = (2m+1)\pi$, with $m$ being an integer). Only when the cells have the same indicatrix orientations ($\varphi_1 = \varphi_2$), the behaviour of the entire sample acquires well familiar features ($\Delta = \Delta_1 + \Delta_2$ and $\beta = 0$).

As mentioned above, the optical anisotropy of photoelastic materials could be regarded as weak enough for the most of practical situations. This implies an extreme smallness of the birefringence. Really, the material is optically isotropic when non-stressed, whereas the amount of anisotropy induced by external (or, all the more, internal) stresses is known to be even much smaller than the (usually very small) natural anisotropy in typical crystalline materials. Nonetheless, it is not quite clear whether the conditions like $\Delta n_l, \Delta n_p \ll 1$ can assure more serious restrictions $\Delta_1,...,\Delta_N \ll 1$, or not. Supposing the cell dimensions to be, in the order of magnitude, as small as a resolved pixel of CCD cameras ($d_1,...,d_N \sim 10 \mu$ m), we arrive at a reasonable estimation $\Delta_1,...,\Delta_N \sim 10^2 \Delta n_l \ll 1$. However, a consistent consideration would then require a further approximation $\Delta \sim N\Delta_1 \ll 1$ or $10^7 d \Delta n_l \ll 1$ for the total retardation, which may be valid for one substances, mechanical stresses and whole sample sizes and invalid for the others. On this basis, we conclude that the validity of the above approximation, which we call as "essentially weak anisotropy" one, is to be checked in each particular case.

If it is indeed fulfilled, radical simplifications follow directly from Eqs. (2)–(7) in the approximation linear in $\Delta_1,...,\Delta_N$. First, the JMs $\mathbf{t}_n$ in Eq. (2) differ only slightly from the unit 2x2-matrix and so commute (see also [10]). This implies that $\mathbf{T_k} \approx \mathbf{T_{-k}}$, together with a practical absence of the "apparent" gyration ($\beta \approx 0$). In other words, the sample becomes a pure linearly birefringent retardation plate, similarly to constituent cells. Besides, from Eq. (5) we obtain

$$\tan 2\varphi \approx \frac{\sum_{i=1}^{N} \Delta_i \sin 2\varphi_i}{\sum_{i=1}^{N} \Delta_i \cos 2\varphi_i}, \qquad (8)$$

i.e. the "effective" optical indicatrix remains arbitrarily aligned with respect to the indicatrices of individual cells. We emphasis at this point that, apart from the evident physical meaning, $\varphi$, $\beta$ and $\Delta$ represent the parameters that can be strictly measured with polarimetric equipment (see [20]). Therefore, experimental verification of the fact $\beta \approx 0$ may serve as a practical criterion of the conditions of "essentially weak anisotropy" for a given sample. Another criterion, coming from the fact $\Delta \ll 1$, is obvious experimental difficulties of finding the optical orientation of sample between the crossed polarizers (see Eq. (8)).

### Relations between the Jones matrix and dielectric parameters

Let us finally concentrate on linking the experimental information about $\varphi$, $\beta$ and $\Delta$ for different directions of probing light beam with the dielectric tensor values $\varepsilon_{ij}$ for each cell. Having fixed the laboratory coordinate system *XYZ*, e.g., to the edges of the cubic sample, one can easily find orientations of the working coordinate systems *xyz* for each propagation direction (see above) and relate the "transverse" dielectric components with those written in the laboratory system ($\varepsilon_{T,ij}$).

As an example, consider the propagation direction parallel to *Z* axis, for which the components $\varepsilon_{11}$, $\varepsilon_{22}$ and $\varepsilon_{12} = \varepsilon_{21}$ are of importance. After diagonalizing $\boldsymbol{\varepsilon}_T$ tensor ($\det(\varepsilon_{ij} - \xi\delta_{ij}) = 0$, with $i,j = 1,2$ and $\delta_{ij}$ being the Kronecker delta), one obtains

$$\xi_{1,2} = [\varepsilon_{11} + \varepsilon_{22} \pm \sqrt{(\varepsilon_{11} - \varepsilon_{22})^2 + 4\varepsilon_{12}^2}]/2, \quad (9)$$

$$\tan 2\varphi = 2\varepsilon_{12}/(\varepsilon_{11} - \varepsilon_{22}), \quad (10)$$

where $\xi_{1,2}$ determine the refractive indices ($n_{1,2} = \sqrt{\xi_{1,2}}$). We remind also that the angle $\varphi$ in Eq. (10) refers to separate cells, not the whole sample (see Eq. (2)). Let us take a smallness of the birefringence $\Delta n = n_1 - n_2$ into consideration. Making use of the conditions $\varepsilon_{11} - \varepsilon_{22}, \varepsilon_{12} \ll \varepsilon_{11}, \varepsilon_{22}$ in Eq. (9) then yields in

$$\Delta n = \sqrt{\left(\sqrt{\varepsilon_{11}} - \sqrt{\varepsilon_{22}}\right)^2 + \varepsilon_{12}^2/\varepsilon}, \quad (11)$$

where the items in the r.h.s. of Eq. (11) determine the linear and "diagonal" linear birefringences for the weak anisotropy case:

$$\Delta n_l = \sqrt{\varepsilon_{11}} - \sqrt{\varepsilon_{22}}, \quad \Delta n_p = \varepsilon_{12}/\sqrt{\varepsilon}. \quad (12)$$

The birefringences in the cells and their indicatrix orientations for the other propagation directions may be calculated in a similar manner. For instance, if the beam propagates along the positive direction bisecting the plane $YZ$, we have

$$\Delta n = \sqrt{\left(\sqrt{\varepsilon_{11}} - \sqrt{(\varepsilon_{22} + \varepsilon_{33})/2} + \varepsilon_{23}/(2\sqrt{\varepsilon})\right)^2 + (\varepsilon_{12} - \varepsilon_{13})^2/(2\varepsilon)}, \quad (13)$$

$$\tan 2\varphi = \frac{\sqrt{2}(\varepsilon_{12} - \varepsilon_{13})}{\varepsilon_{11} - (\varepsilon_{22} + \varepsilon_{33})/2 + \varepsilon_{23}}, \quad \text{an so on.} \quad (14)$$

A more general method may be used when finding the JM **t** in terms of $\boldsymbol{\varepsilon}_T$, based on the relations between **t** and the corresponding differential JM **M** [10], which, in its turn, depends upon $\boldsymbol{\varepsilon}_T$. The method employs the fact that the eigenvectors of **t** and **M** are the same, while for the eigenvalues $\gamma$ one has $\gamma_{\mathbf{t}} = \exp[i(\pi d/\lambda)\gamma_{\mathbf{M}}]$ (see, e.g., [2,20]). However, it does not account consistently a weakness of optical anisotropy, the point that allows for utilising the JM approach itself. Notice also that the particular relationships between **t** and the dielectric parameters available in [2,20] include some inaccuracies.

Collecting the experimental $\varphi$, $\beta$ and $\Delta$ data for a sufficiently large number of propagation directions [*] and using the formulae given above would enable determining $\varepsilon_{ij}$ for all of the cells. At the same time, it is the easiest to find the parameter $\varepsilon$ from any measurements of the mean refractive index.

## Conclusions

In the present work we have related the problem of crystal optical properties of inhomogeneous dielectric materials with the issues of 3D photoelasticity and polarization-optical tomography of 3D tensor fields. The approach adopted here is description in terms of integral JMs. A major attention is given to a specific case of "discretely inhomogeneous" composite crystalline plates, concerned with the problems of the cell tomographic model. Some ideas of integrated photoelasticity are reformulated and thoroughly substantiated. The approach of uniform "effective" optical medium is developed. We suggest describing polarization effect of the material, which is inhomogeneous along the propagation direction of the light beam, in terms of the "effective" phase retardation, optical activity and indicatrix orientation. They are equivalent to the earlier known characteristic azimuths and characteristic retardation, but bear a more evident physical meaning.

---

[*] The number of probing beam directions mentioned in [20] is to be qualified as a minimum, necessary for unambiguous reconstruction of $6N^3$ components of the symmetric dielectric tensor. However, practical considerations of accuracy and numerical stability of solutions of the inverse tomographic problem would, of course, require much more experimental data.

We have shown that a spatial dependence of optical indicatrix orientation results in an "apparent" gyration effect. The other unusual result is "non-additive" total phase retardation of the inhomogeneous material, or a necessity in accounting for the indicatrix orientations when finding this retardation. Different practical approximations employed in frame of polarimetric tomography are critically discussed, in particular those of "weak" and "essentially weak" optical anisotropy. Practical formulae relating the JM of photoelastic material and its dielectric tensor are obtained for these limiting cases. Further consideration of the mentioned points, as well as a comparison with experimental data, will be a subject of forthcoming paper.

**Acknowledgement**

The authors acknowledge financial support from the Scientific and Technology Centre in Ukraine (Project No 3042).